\documentclass[12pt]{article}
\usepackage{epsfig,amssymb,psfrag}

\catcode`\@=11
\def \thesection {\arabic{section}.}

\def \be  {\begin{equation}}
\def \ee  {\end{equation}}
\def \ba  {\begin{eqnarray}}
\def \ea  {\end{eqnarray}}
\def \baa {\begin{eqnarray*}}
\def \eaa {\end{eqnarray*}}
\def \bb  {\begin {thebibliography} }
\def \eb  {\end{thebibliography}}
\def \lab #1 {\label{#1}}

\newcommand \bi [1] {\bibitem{#1}}
\newcommand\re[1]{(\ref{#1})}

\def \matrix #1 {\left(\begin{array}{cc} #1 \end{array}\right)}

\def \Im {\mathop{\rm Im}\nolimits}
\def \Re {\mathop{\rm Re}\nolimits}

\newcommand\lr[1]{{\left({#1}\right)}}
\newcommand \widebar [1] {\overline{#1}}

\newcommand \vev [1] {\langle{#1}\rangle}

\newcommand{\as}{\ifmmode\alpha_{\rm s}\else{$\alpha_{\rm s}$}\fi}
\newcommand{\asbar}{\ifmmode\bar{\alpha}_{\rm s}\else{$\bar{\alpha}_{\rm s}$}\fi}

\font\cmss=cmss12 
\def\inbar{\,\vrule height1.5ex width.4pt depth0pt}
\def\IC{\relax\hbox{$\inbar\kern-.3em{\rm C}$}}
\def\IZ{\relax{\hbox{\cmss Z\kern-.4em Z}}}
\def\IR{{\hbox{{\rm I}\kern-.2em\hbox{\rm R}}}}

\def\IP{{\hbox{{\rm I}\kern-.2em\hbox{\rm P}}}}
\def\II{\hbox{{1}\kern-.25em\hbox{l}}}

\def\numberbysection{\@addtoreset{equation}{section}
                     \def\theequation{\thesection\arabic{equation}}}
\numberbysection

\newcommand \Mybf[1] {\mbox{\boldmath$ {#1} $}}
\newcommand \mybf[1] {\mbox{\boldmath$ {\scriptstyle #1} $}}

\begin{document}

\begin{titlepage}
\begin{flushright}
\begin{tabular}{l}
LPT--Orsay--03--37\\
TPJU-2/2003 \\
RUB--TP2--08/03\\
hep-ph/0306250
\end{tabular}
\end{flushright}

\vskip3cm
\begin{center}
  {\large \bf
Multi-reggeon compound states and
 \\[2mm] resummed anomalous dimensions in QCD}

\def\thefootnote{\fnsymbol{footnote}}%
\vspace{1cm}
{\sc G.P.~Korchemsky}${}^1$, {\sc J.~Kota{\'n}ski}${}^2$ and {\sc
A.N.~Manashov}${}^3$\footnote{ Permanent address:\ Department of Theoretical
Physics,  Sankt-Petersburg State University, St.-Petersburg, Russia}
\\[0.5cm]

\vspace*{0.1cm} ${}^1$ {\it
Laboratoire de Physique Th\'eorique
\footnote{Unite Mixte de Recherche du CNRS (UMR 8627)},
Universit\'e de Paris XI, \\
91405 Orsay C\'edex, France
                       } \\[0.2cm]
\vspace*{0.1cm} ${}^2$ {\it
Institute of Physics, Jagellonian University,\\ Reymonta 4, PL-30-059 Cracow,
Poland
                       } \\[0.2cm]
\vspace*{0.1cm} ${}^3$
{\it Institut f\"ur Theoretische Physik II, Ruhr--Universit\"at Bochum, \\
 44780 Bochum, Germany}

\vskip2cm
{\bf Abstract:\\[10pt]} \parbox[t]{\textwidth}{We perform the OPE analysis of
the contribution of colour-singlet compound states of reggeized gluons to a
generic hard process in QCD and calculate the spectrum of the corresponding
higher twist anomalous dimensions in multi-colour limit. These states govern high
energy asymptotics of the structure functions and their energies define the
intercept of the Regge singularities both in the Pomeron and the Odderon sectors.
We argue that due to nontrivial analytical properties of the energy spectrum, the
twist expansion does not hold for the gluonic states with the minimal energy
generating the leading Regge singularities. It is restored however after one
takes into account the states with larger energies whose contribution to the
Regge asymptotics is subleading.}

\vskip1cm

\end{center}

\end{titlepage}



\newpage

\setcounter{footnote} 0

\section{Introduction}

Scale dependence of hadronic cross-sections is driven by anomalous dimensions
calculable in perturbative QCD as series in the coupling constant. The classical
example is provided by the deeply inelastic scattering of a virtual photon
$\gamma^*(q)$ off a (polarized) hadron with momentum $p_\mu$. In that case, the
operator product expansion (OPE) allows one to expand the moments of the
structure function $F(x,Q^2)$ (with $x=Q^2/2(p\cdot q)$ and $Q^2=-q_\mu^2$) in
inverse powers of hard scale $Q$ and identify the expansion coefficients as the
forward matrix elements of Wilson operators $\mathcal{O}^a_{n,j}(0)$ of
increasing twist $n\ge 2$ and Lorentz spin $(j-1)\ge 1$
\be
\widetilde F(j,Q^2)\equiv \int_0^1 dx\, x^{j-2} F(x,Q^2)=\sum_{n=2}^\infty
\frac1{Q^n} \sum_{a} C_n^a(j,\alpha_s(Q^2))\, \vev{p\,|\mathcal{O}^a_{n,j}(0)|p}\,,
\label{moments}
\ee
where $C_n^a(j,\alpha_s)$ are the corresponding coefficient functions and the
superscript $a$ is introduced to enumerate operators of the same twist $n$. Their
total number depends on $n$ and rapidly grows for $n\ge 3$. The $Q^2-$dependence
of $\widetilde F(j,Q^2)$ follows from the dependence of the twist$-n$ operators
on the normalization scale $\mu^2=Q^2$. In general, these operators mix under
renormalization. Diagonalizing the corresponding mixing matrix, one can construct
multiplicatively renormalizable operators
\be
 Q^2\frac{d}{d Q^2}\vev{p\,|\mathcal{O}^a_{n,j}(0)|p}
=\gamma_n^a(j)\,\vev{p\,|\mathcal{O}^a_{n,j}(0)|p}\,,
\label{anom-dim}
\ee
where the anomalous dimension has a perturbative expansion
$\gamma_n^a(j)=\sum_{k=1}^\infty\gamma_{k,n}^a(j) (\alpha_s(Q^2)/\pi)^k $ with
the expansion coefficients $\gamma_{k,n}^a(j)$ having a nontrivial
$j-$dependence.


For $j\to 1$, the moments \re{moments} receive the dominant contribution from the
small$-x$ region, in which the structure function has a Regge behaviour
$F(x,Q^2)\sim (1/x)^{\alpha-1}$, with $\alpha$ close to unity. To find the scale
dependence of the structure function $F(x,Q^2)$ at small $x$, one has to
analytically continue the anomalous dimensions $\gamma_n^a(j)$ from positive
integer $j\ge 2$ to ``unphysical'' $j\sim 1$ and invert the moments \re{moments}.
For twist two this can be done using the well-known DGLAP expressions for
$\gamma_{n=2}^a(j)$. For twist $n\ge 3$, the calculation of $\gamma_n^a(j)$ is
much more involved already for positive integer $j\ge 2$ mainly because the
number of operators increases with the twist and the size of the corresponding
mixing matrices depends on $j$. As a consequence, analytical continuation of
anomalous dimensions of the operators of twist $n\ge 3$ within the conventional
OPE approach turns out to be an extremely difficult (if not impossible) task.

Another approach to finding the asymptotic behaviour of the anomalous dimensions
$\gamma_n^a(j)$ for $j\to 1$ has been proposed in Ref.~\cite{T}. It relies on the
relation between the twist expansion of the moments \re{moments} and the
small$-x$ behaviour of the structure functions $F(x,Q^2)$ obtained within the
framework of the BFKL approach~\cite{BFKL}. Since the anomalous dimensions do not
depend on the choice of the scattered particles, one can simplify the analysis by
considering the deeply inelastic scattering of a virtual photon $\gamma^*(Q^2)$
off a perturbative ``onium'' state of the mass $p_\mu^2=M^2$, such that
$\Lambda_{\rm QCD}^2\ll M^2\ll Q^2$. In the BFKL approach, one obtains the
small$-x$ behaviour in this process by resumming a special class of perturbative
corrections enhanced by powers of $\alpha_s\ln(1/x)$. Among them there are
$\alpha_s\ln(1/x)\ln (Q^2/M^2)-$terms responsible for anomalous $Q^2-$dependence
of $F(x,Q^2)$. Expanding the resulting expression for the moments $\widetilde
F(j,Q^2)$ in powers of $1/Q$, one can separate the twist$-n$ contribution and
calculate the corresponding anomalous dimensions $\gamma_n^a(j)$ for $j\to
1$\footnote{It is known \cite{Mueller1996}, that the OPE expansion breaks down
for very small values of $x$, away from the region
$\ln(1/x)\ll\ln^2(Q^2/\Lambda_{\rm QCD}^2)$. We shall assume that $Q^2$ is large
enough so that this condition is fulfilled.}. Still, the actual form of the
underlying twist$-n$ operators remains to be found. It has been conjectured
\cite{L85,DL} that they belong to the class of quasipartonic operators.

To begin with, we recall that in the BFKL approach, in the generalized leading
logarithmic approximation (GLLA) \cite{BKP}, the structure function takes the
following form at small$-x$
\\[1.5mm]
\be
F(x,Q^2)= \sum_{N\ge 2}\bar\alpha_s^{N-2} F_N(x,Q^2)\,,\qquad
F_N(x,Q^2)=\sum_{\mybf{q}} (1/x)^{-\bar \alpha_s E_N(\mybf{q})}
\beta^{\,\mybf{q}}_{\gamma^*}(Q)\beta^{\,\mybf{q}}_p(M)\,,
\label{GLLA}
\ee
with $\bar\alpha_s=\alpha_sN_c/\pi$. Here  $F_N(x,Q^2)$ describes the
contribution of color-singlet compound states built from $N=2,3,\ldots$ reggeized
gluons, or briefly the $N-$reggeon states. These states
satisfy a Schr\"odinger-like equation for the system of $N$ gluons interacting on
the two-dimensional plane of transverse coordinates \cite{BKP}. Their spectrum is
labelled in \re{GLLA} by the set of quantum numbers $\Mybf{q}$. For $x\to 0$, the
sum over $\Mybf{q}$ in \re{GLLA} is dominated by the state with the minimal
energy $E_N(\Mybf{q})$. The contribution of this state to $F_N(x,Q^2)$ has the
Regge form $F_N(x,Q^2)\sim (1/x)^{-\bar\alpha_s E_N(\mybf{q})}$ and its
$Q-$dependence is carried by the residue factors
$\beta^{\,\mybf{q}}_{\gamma^*}(Q)\beta^{\,\mybf{q}}_p(M)$ which measure the
overlap of the wave function of the state with the wave functions of scattered
particles.

For $\bar\alpha_s\ln x\sim 1$ and $x\to 0$, the sum over $N$ in \re{GLLA} is
dominated by the $N=2$ term and gives rise to the BFKL Pomeron~\cite{BFKL}.
%
Expanding its contribution to \re{moments} in powers of $1/Q$, one finds that the
$N=2$ reggeon states generate an infinite tower of composite gluonic operators of
increasing twist $n\ge 2$ \cite{L85}. At $n=2$, the anomalous dimensions of
twist-two (gluonic) operators, $\gamma_2(j)$, satisfy in the leading logarithmic
approximation, $\bar\alpha_s/(j-1)=\rm fixed$ and $j\to 1$, the master
equation~\cite{T}
\be
\lr{\frac{\bar\alpha_s}{j-1}}^{-1}=2\psi(1)-\psi(\gamma_2(j))
-\psi(1-\gamma_2(j))\,,
\label{master-eq}
\ee
where 
$\psi(x)=d\ln\Gamma(x)/dx$ is the Euler digamma function. Its solution looks like
\be
\gamma_2(j)=\frac{\bar\alpha_s}{j-1}+2\zeta(3)\lr{\frac{\bar\alpha_s}{j-1}}^4
+2\zeta(6)\lr{\frac{\bar\alpha_s}{j-1}}^6 +\mathcal{O}(\bar\alpha_s^8)\,,
\label{sol-2}
\ee
with $\zeta(k)$ the Riemann zeta-function. Eq.\,\re{sol-2} is in agreement with
the well-known two-loop expression for the twist-two anomalous dimension
$\gamma^{\rm GG}_j$. The series \re{sol-2} diverges for
$j-1=\mathcal{O}(\alpha_s)$ and its radius of convergence is determined by the
right-most Regge singularity, $\gamma_2(j)\sim \sqrt{j-j_2}$ with $j_2=1+
4\widebar\alpha_s\ln 2$ being the intercept of the BFKL Pomeron~\cite{BFKL}.

The contribution of the $N\ge 3$ reggeon states to \re{GLLA} is suppressed by a
power of the coupling constant and it serves to unitarize the asymptotic
behaviour of the structure function $F(x,Q^2)$ at very high energy. To perform
the OPE analysis of these states, one has to expand the moments of $F_N(x,Q^2)$
in inverse powers of $Q$ and match the resulting expression for $\widetilde
F_N(j,Q^2)$ into \re{moments}.
Similar to the $N=2$ case, the resummed anomalous dimensions of these operators
can be deduced from the spectrum of the underlying $N-$reggeon states.  This
program has been carried out in Refs.~\cite{T,L85} only for the $N=2$ states~
while for $N\ge 3$ the main difficulty was a poor understanding of the properties
of the $N-$reggeon states. Later, the anomalous dimensions of twist-four gluonic
operators have been calculated in the double-logarithmic approximation in
Refs.~\cite{LRS,B}. These studies revealed that due to the presence of Regge
cuts, the anomalous dimension of $N-$reggeon states has a complicated analytical
structure which get simplified however in the multi-color limit. Inspired by this
observation, we perform in the present paper the OPE analysis to the $N\ge 3$
compound states in the limit $N_c\to\infty$. Namely, we identify the leading
twist contribution of these states to the moments \re{moments} and calculate the
corresponding anomalous dimensions for $j\to 1$ in the multi-color limit. Our
analysis relies on the remarkable integrability properties of the $N-$reggeon
states \cite{L1,FK} and on the exact expressions for the energy spectrum of these
states found in \cite{DKKM1,DKKM}.

The paper is organized as follows. In Sect.~2 we summarize the properties of the
multi-reggeon states, perform the OPE expansion of their contribution to the
structure function and establish the master equation for anomalous dimensions of
high twist operators. This equation involves analytical continuation of the
energy spectrum of the $N-$reggeon states in multi-colour QCD, which is performed
in Sect.~3. In Sect.~4 we present our results for the anomalous dimensions.
Sect.~5 contains concluding remarks.

\section{Reggeon compound states in multi-colour QCD}

Let us perform the OPE expansion of the contribution of the $N$ reggeon compound
state to the structure function \re{moments}. Going over to the moment space we
find from \re{GLLA}
\be
\widetilde F_N(j,Q^2)
=\sum_{\mybf{q}} \frac1{j-1+\bar \alpha_s E_N(\Mybf{q})}
\beta_{\gamma^*}^{\,\mybf{q}}(Q)\,
 \beta_p^{\,\mybf{q}}(M) \,,
\label{N-moment}
\ee
where the sum runs over the spectrum of the $N-$reggeon states. These states
satisfy the Schr\"odinger equation \cite{BKP} which possesses, in the multi-color
limit, ``hidden'' conserved charges $\Mybf{q}$ and, as a consequence, is
completely integrable~\cite{L1,FK}.

Due to complete integrability, the spectrum of the $N-$reggeon states in
multi-colour QCD is uniquely specified by a complete set of the quantum numbers
$\Mybf{q}=(q_2,\bar q_2\ldots,q_N,\bar q_N)$ defined as eigenvalues of the
corresponding integrals of motion. The wave function of the state,
$\Psi_{\mybf{q}}(\{\vec z\};\vec z_0)$, depends on two-dimensional impact
parameters of $N$ reggeons, $\{\vec z\}\equiv(\vec z_1,\ldots,\vec z_N)$, and the
impact parameter of the center-of-mass of the state, $\vec z_0$. Introducing
auxiliary complex (anti)holomorphic coordinates, $z=x+iy$ and $\bar z=x-iy$, one
finds \cite{L85} that the effective QCD Hamiltonian for the system of
$N-$reggeons is invariant under conformal, $SL(2,\mathbb{C})$ transformations on
the plane $z\to (az+b)/(cz+d)$ and $\bar z\to (\bar a\bar z+\bar b)/(\bar c\bar
z+\bar d)$ with $ad-bc=\bar a\bar d-\bar b\bar c=1$. As a consequence, its
eigenstates $\Psi_{\mybf{q}}(\{\vec z\};\vec z_0)$ belong to irreducible
representation of
the $SL(2,\mathbb{C})$ group:
\be
\Psi_{\mybf{q}}(\{\vec z\};\vec z_0)\to (cz_0+d)^{2h}(\bar c \bar z_0+\bar
d)^{2\bar h} \prod_{k=1}^N (cz_k+d)^{2s}(\bar c \bar z_k+\bar d)^{2\bar
s}\Psi_{\mybf{q}}(\{\vec z\};\vec z_0)\,,
\label{SL2}
\ee
where $(s=0,\bar s=1)$ is the $SL(2,\mathbb{C})$ spin of a single reggeon and the
parameters $(h,\bar h)$ define the $SL(2,\mathbb{C})$ spin of the $N-$reggeon
state. For the principal series of the $SL(2,\mathbb{C})$, their possible values
are parameterized by nonnegative integer $n_h$ and real $\nu_h$
\be
h=\frac{1+n_h}2+i\nu_h\,,\qquad \bar h=\frac{1-n_h}2+i\nu_h\,.
\label{h}
\ee
By the definition, the wave function $\Psi_{\mybf{q}}(\{\vec z\};\vec z_0)$ has
to diagonalize the integrals of motion in the $z-$ and $\bar z-$sectors,
$\Mybf{q}=(q_2,\bar q_2\ldots,q_N,\bar q_N)$. Also, it has to be a single-valued
function on the two-dimensional $\vec z-$plane and be normalizable with respect
to the $SL(2,\mathbb{C})$ scalar product. These requirements lead to the
quantization conditions for the integrals of motion.

As was shown in \cite{DKKM}, the quantized values of $\Mybf{q}$ depend on real
$\nu_h$ and integer $n_h\ge 0$ defined in \re{h} and on the set of integers
$\Mybf{\ell}=(\ell_1,\ldots,\ell_{2N-4})$. Then, the sum over the $N-$reggeon
states in \re{N-moment} looks like $\sum_{\mybf{q}}\equiv
\sum_{\mybf{\ell}}\int_{-\infty}^\infty d\nu_h \sum_{n_h=0}^\infty$. The
eigenstates with different quantum numbers
$\Mybf{q}=\Mybf{q}(\nu_h;n_h,\Mybf{\ell})$ are orthogonal 
with respect to the $SL(2,\mathbb{C})$ invariant scalar product
\\[-3mm]
\be
\vev{\Psi_{\mybf{q}}(\vec z_0)|\Psi_{\mybf{q}'}(\vec z_0')}\equiv\int
\prod_{k=1}^N d^2 z_k 
\Psi_{\mybf{q}}(\{\vec z\};\vec z_0) \lr{\Psi_{\mybf{q}'}(\{\vec z\};\vec
z_0')}^*= \delta^{(2)}(z_0-z_0')\,\delta_{\mybf{q}\mybf{q}'}\,,
\label{norm}
\ee
\\[-2mm]
where $\delta_{\mybf{q}\mybf{q}'}\equiv \delta(\nu_h-\nu_h')
\delta_{n_hn_h'}\delta_{\ell\ell'}$. The energy of the $N-$reggeon state,
$E_N=E_N(\nu_h;n_h,\Mybf{\ell})$, is a real continuous function of $\nu_h$. For
different integer $n_h$ and $\Mybf{\ell}$, these functions form an infinite set
of ``trajectories'' (see Fig.~1 on the left).

As we will see in a moment, to calculate the anomalous dimensions we will have to
perform an analytical continuation of the energy spectrum of the $N-$reggeon
state from ``physical'' values of the conformal spin $h$, Eq.~\re{h}, to
arbitrary complex $h$. Since $h$ depends on two parameters, $n_h$ and $\nu_h$,
one has to decide which of these parameters (if not both) can be made complex. To
this end, we note that $n_h$ and $\nu_h$ have a different physical meaning.
According to \re{SL2}, they define the two-dimensional Lorentz spin, $h-\bar
h=n_h$, and the scaling dimension, $h+\bar h=1+2i\nu_h$, of the $N-$reggeon
state, respectively. For the wave function $\Psi_{\mybf{q}}(\{\vec z\};\vec z_0)$
to be single-valued, $n_h$ has to be integer while reality condition for $\nu_h$
follows from the requirement that $\Psi_{\mybf{q}}(\{\vec z\};\vec z_0)$ has to
be normalizable with respect to the scalar product \re{norm}. Performing
analytical continuation, we shall require that $\Psi_{\mybf{q}}(\{\vec z\};\vec
z_0)$ has to be single-valued on the $\vec z-$plane for arbitrary complex $h$ and
lift the normalizability condition. This implies that $n_h$ has to be integer and
analytical continuation goes in $\nu_h$. The function $\Psi_{\mybf{q}}(\{\vec
z\};\vec z_0)$ defined in this way obeys \re{SL2}, diagonalizes the integrals of
motion $\Mybf{q}$ but it does not satisfy \re{norm} for arbitrary complex $h$.

To expand \re{N-moment} in inverse powers of $Q$, one examines the expression for
the residue factors. They describe the coupling of the $N-$reggeon states to the
scattered particles and are given by
\be
\beta^{\mybf{q}}_{\gamma^*}(Q)=\int d^2 z_0\, \vev{\Psi_{\gamma^*}|\Psi_{\mybf{q}}(\vec z_0)}\,,
\qquad
\beta^{\mybf{q}}_p(M)=\int d^2 z_0\, \vev{\Psi_{\mybf{q}}(\vec z_0)|\Psi_p}\,.
\label{imp-fac}
\ee
Here the scalar product is taken with respect to \re{norm} while
$\Psi_{\gamma^*}(\{\vec z\})$ and $\Psi_p(\{\vec z\})$ are the wave functions of
the photon and the onium state, respectively, in the impact parameter
representation. The $\vec z_0-$integration in \re{imp-fac} ensures that the total
momentum transferred in the $t-$channel equals zero. Neglecting the running of
the coupling constant, one finds that the impact factors depend on a single scale
-- the invariant mass of the particle. From dimensional counting, the dependence
is fixed by the scaling dimension of the $N-$reggeon state
\be
\beta^{\mybf{q}}_{\gamma^*}(Q)= C_{\gamma^*}^{\mybf{q}} Q^{-1-2i\nu_h}\,,\qquad
\beta^{\mybf{q}}_p(M)=C_{p}^{\mybf{q}}\, M^{-1+2i\nu_h}\,,
\label{sc-imp}
\ee
where $C_{\gamma^*}^{\mybf{q}}$ and $C_{p}^{\mybf{q}}$ are dimensionless
coefficients depending on the charges $\Mybf{q}=\Mybf{q}(\nu_h;n_h,\Mybf{\ell})$.
They are different from zero provided that the reggeon state has the same quantum
numbers (Bose symmetry,  C-parity, two-dimensional angular momentum) as hadronic
states to which it couples. In particular, $C_{\gamma^*}^{\mybf{q}}$ vanishes for
the reggeon states with the Odderon quantum numbers. Our subsequent consideration
does not rely on the properties of $C_{\gamma^*}^{\mybf{q}}$ and it is valid for
a generic short-distance dominated process which receives a nonvanishing
contribution from the reggeon states both in the Odderon and the Pomeron sectors.

Substituting \re{sc-imp} into \re{N-moment}, 
one gets
\be
\widetilde F_N(j,Q^2)=\frac{1}{Q^2}\sum_{\mybf{\ell}}\sum_{n_h\ge 0 }
\int_{-\infty}^\infty d\nu_h
\frac{C_{\gamma^*}^{\mybf{q}}\,C_{p^{\phantom{*}}}^{\mybf{q}}}{j-1+\bar \alpha_s
E_N(\Mybf{q})} \lr{\frac{M}{Q}}^{-1+2i\nu_h}\,.
\label{F-fin}
\ee
Let us examine \re{F-fin} in two limits: (i) $Q^2/M^2=\rm fixed$, $j\to 1$ and
(ii) $Q^2/M^2\to\infty$, $j=\rm fixed$.

In the limit (i), one obtains from \re{F-fin} that the rightmost Regge
singularity of $\widetilde F_N(j,Q^2)$ is located at $j_N=1 - {\widebar \alpha_s}
\min{}_{q}E_N(\Mybf{q})$
\be
\widetilde F_N^{\rm(i)}(j,Q^2) \sim \frac1{Q^2} \int \frac
{d\nu_h}{j-j_N+\sigma_N\nu_h^2} \lr{\frac{M}{Q}}^{-1+2i\nu_h} \sim
\frac1{Q}\frac1{\sqrt{j-j_N}}\,.
\label{sq-root}
\ee
Here the dispersion parameter $\sigma_N$ describes the $\nu_h-$dependence of the
energy $E_N(\Mybf{q})$ around its minimal value. Eq.~\re{sq-root} leads to a
power rise of the structure function at small$-x$, $F_N(x,Q^2)\sim Q^{-1}\,
(1/x)^{j_N}/(\ln 1/x)^{1/2}$, in agreement with the properties of the BFKL
Pomeron ($N=2$) \cite{BFKL} and higher $N\ge 3$ reggeon states~\cite{DKKM},

In the limit (ii), we choose $j> j_N$ and expand the r.h.s.\ of \re{F-fin} in
powers of $M/Q$. For $M/Q\to 0$, the $\nu_h-$integration in \re{F-fin} can be
performed by deforming the integration contour into the lower half-plane and
picking up the contribution from singularities of the integrand in \re{F-fin}.
The latter could come from singularities of the impact factors
$C_{\gamma^*}^{\mybf{q}}\,C_{p^{\phantom{*}}}^{\mybf{q}}$, possible branch cuts
of the energy $E_N(\Mybf{q})$ and zeros of the denominator. Let us examine these
three possibilities one after another.

It is known \cite{T} that the impact factors may have poles in $\nu_h$ due to
mixing between gluonic and quark operators. Since the mixing does not affect the
leading $j\to 1$ asymptotics of gluonic operators, we can safely neglect it.
Next, one has to examine the possibility that the energy $E_N(\Mybf{q})$ has cuts
on the complex $\nu_h-$plane \cite{LRS,B}. In that case, as we will show below,
the integration around the cut in \re{F-fin} provides a nontrivial contribution
to $\widetilde F_N(j,Q^2)$ which contradicts \re{moments}. For the OPE expansion
\re{moments} to be valid, the contribution of cuts should cancel in the sum
\re{F-fin} over the $N-$reggeon states. We will argue in Sect.~4 that this is
exactly what happens for $N\ge 3$ reggeon states. Finally, the poles of
\re{F-fin} 
originating from zeros of the denominator can be defined as solutions to the
master equation
\be
j-1+\bar\alpha_s E_N(\Mybf{q})=0\,,
\label{ME}
\ee
where $\Mybf{q}=\Mybf{q}(\nu_h;n_h,\Mybf{\ell})$. At $N=2$ this equation matches
\re{master-eq} for $\gamma_2(j)=1/2-i\nu_h$. For $N\ge 3$, the general solution
to \re{ME} takes the form $\nu_h=\nu_h(\widebar\alpha_s/(j-1);n_h,\Mybf{\ell})$.
As follows from \re{F-fin}, its contribution to the moments of structure function
scales as $\widetilde F_N(j,Q^2)\sim Q^{-2} (M/Q)^{-1+2i\nu_h}$. It matches the
OPE expansion, Eqs.~\re{moments} and \re{anom-dim}, provided that
$n-2\gamma_n(j)=1+2i\nu_h$, or equivalently
\be
\gamma_n(j)=(n-1)/2-i\nu_h=[n-(h+\bar h)]/2\,,
\label{expect1}
\ee
where $h$ and $\bar h$ are the $SL(2,\mathbb{C})$ spins defined in \re{h}. This
equation establishes the relation between the anomalous dimension of the
twist$-n$ operator $\gamma_n(j)$, Eq.~\re{anom-dim}, and solutions to \re{ME}.
Since this operator is built from $N$ gluon strength tensors and carries the
two-dimensional Lorentz spin $n_h\ge 0$, its twists satisfies $n\ge N+n_h$. In
addition, writing the anomalous dimension as
$\gamma_n(j)=\gamma_{n}^{(0)}{\bar\alpha_s}/(j-1)+{\cal O}(\bar\alpha_s^2)$, one
expects that solutions to \re{ME} should look like
$i\nu_h=(n-1)/2-\gamma_{n}^{(0)}{\bar\alpha_s}/(j-1)+{\cal O}(\bar\alpha_s^2)$
with $n\ge N+n_h$. Combining this expression together with \re{ME} one finds
\be
E_N(\Mybf{q})\sim \frac{\gamma_{n}^{(0)}}{i\nu_h-(n-1)/2}\,.
\label{expect}
\ee
Thus, in order for the moments \re{F-fin} to admit the OPE expansion
\re{moments}, the energy of the $N-$reggeon state has to contain an infinite set
of poles located along imaginary $\nu_h-$axis at (half)integer points
$\nu_h=-i(n-1)/2$ with $n\ge N+n_h$. For $i\nu_h=(n-1)/2+\epsilon$ the Laurent
expansion of the energy around the pole at $\epsilon=0$ can be written as
\be
E_N(\Mybf{q})={\frac{c_{-1}}{\epsilon}-c_0-c_1\,\epsilon+\ldots}\,.
\label{E-pole-exp}
\ee
Its substitution into \re{ME} yields the following expression for the anomalous
dimension of the twist$-n$ operators, $\gamma_n(j)=-\epsilon$,
\be
\gamma_n(j)=c_{{-1}}\left[\frac{\bar\alpha_s}{\omega}+c_{{0}}\,
\lr{\frac{\bar\alpha_s}{\omega}}^2+\left( c_{0}^{\,2}-c_{{1}}c_{{-1}}
\right)\lr{\frac{\bar\alpha_s}{\omega}}^3+\ldots\right].
\label{gamma-pole-exp}
\ee
with $\omega=j-1$. The coefficients $c_k$ depend on the twist of the operator and
on the quantum numbers specifying the $N-$reggeon state,
$c_k=c_k(n,n_h,\Mybf{\ell})$. The $\mathcal{O}(\alpha_s)-$term in the r.h.s.\ of
\re{gamma-pole-exp} defines the anomalous dimension in the double-logarithmic
approximation~\cite{LRS,B,S} and leads to the following scaling behaviour of the
structure function \re{moments} at small$-x$
\be
F^{\rm(ii)}(x,Q^2) \sim \frac1{Q^n} \exp\lr{2\sqrt{c_{-1} \ln(1/x)\ln Q^2}}\,.
\ee
We conclude that in order to calculate the twist$-n$ anomalous dimensions
\re{gamma-pole-exp} one has to analytically continue the energy of the
$N-$reggeon compound states $E_N(\Mybf{q})$ from ``physical'' values of the
conformal $SL(2,\mathbb{C})$ spins \re{h} to the complex $\nu_h-$plane and
calculate the coefficients of its Laurent expansion around the poles located at
$i\nu_h=(n-1)/2$ with $n\ge N+n_h$, or equivalently $h=(n+n_h)/2$ and $\bar
h=(n-n_h)/2$.

\section{Analytical continuation}

The $N-$reggeon states are the eigenstates of the effective QCD Hamiltonian,
which is a Hermitian operator on the space of functions satisfying \re{SL2},
\re{h} and \re{norm} with $\nu_h$ real and $n_h$ integer. The energies of these
states, $E_N(\Mybf{q})$, are smooth real functions on the real $\nu_h-$axis and
their wave functions $\Psi_{\mybf{q}}(\{\vec z\};\vec z_0)$ are orthogonal to
each other with respect to the scalar product \re{norm}. Performing analytical
continuation of the energy spectrum 
from the real $\nu_h-$axis to the complex $\nu_h-$plane, one
replaces the normalization condition \re{norm}
by a weaker condition for $\Psi_{\mybf{q}}(\{\vec z\};\vec z_0)$ to be a
single-valued function on the
$\vec z-$plane. 
This condition ensures that the two-dimensional integrals entering \re{imp-fac}
are well-defined for complex $\nu_h$.

{}From point of view of quantum mechanics, the problem amounts to finding
analytical properties of the energy $E=E(g)$ as a function of the coupling
constant $g=\nu_h$. In past, this fundamental problem has been thoroughly studied
in various models~\cite{BW,TU,BT,KN}. It was found that, in general, $E(g)$ is a
{\it multi-valued\/} function of complex $g$ and the number of its different
branches is equal to the number of the energy levels at real $g$ when the
Hamiltonian is hermitian. To study a global behaviour of the function $E(g)$, it
proves convenient to glue together different ``sheets'' corresponding to its
branches and consider $E(g)$ as a single-valued function on the resulting Riemann
surface. Depending on the model, this surface may have a rather complicated
structure and consist of a few disconnected parts due to some additional
symmetry~\cite{TU}. A remarkable property of the Riemann surface is that it
encodes the {\it entire\/} spectrum of the model. Namely, knowing the energy of
the ground state at real $g$, one can reconstruct the whole spectrum of the model
by going around the branching points to other sheets of the Riemann surface
corresponding to excited energy levels. We shall demonstrate below that upon
analytical continuation to the complex $\nu_h$ the energy of the $N-$reggeon
compound states $E_N$ shares the same properties (for $N\ge 3$).

To begin with, let us consider the well-known expression for the energy of the
$N=2$ states~\cite{BFKL}
\be
E_2(\nu_h,n_h)=\psi\lr{\frac{1+n_h}2+i\nu_h}
+\psi\lr{\frac{1+n_h}2-i\nu_h}-2\psi(1)\,,
\label{E2-anal}
\ee
with $n_h$ nonnegative integer. $E_2(\nu_h,n_h)$ is a smooth even function on the
real $\nu_h-$axis. It takes its minimal value at $\nu_h=0$ and increases
monotonically for $\nu_h\to\pm\infty$. After analytical continuation,
$E_2(\nu_h,n_h)$ becomes a meromorphic function of $\nu_h$. It has an infinite
set of poles located along imaginary $\nu_h-$axis at $i\nu_h=\pm(n-1)/2$ with
$n\ge 2+n_h$.
The leading twist contribution, $n=2$, corresponds to $n_h=0$. To obtain the
anomalous dimension of twist two, Eq.~\re{sol-2}, one matches the Laurent
expansion of $E_2(\nu_h,0)$ around the pole at $\nu_h=-i/2$ into \re{E-pole-exp}
and applies \re{gamma-pole-exp}. The BFKL Pomeron is located on the same complex
curve $E_2(\nu_h,0)$ at $\nu_h=0$ so that $E_2(0,0)=-4\ln 2$. For $n_h\ge 1$ the
complex curve $E_2(\nu_h,n_h)$ describes subleading Regge singularities and, at
the same time, it generates contribution of higher twist.

For $N\ge 3$ reggeon states, analytical continuation of the energy spectrum is
more subtle. Firstly, the energy of the $N-$reggeon states does not admit a
simple representation like \re{E2-anal} and, secondly,
$E_N=E_N(\nu_h;n_h,\Mybf{\ell})$ depends on the set of $2(N-2)$ integers
$\Mybf{\ell}$, which parameterize eigenvalues of the integrals of motion. Thus,
performing analytical continuation of $E_N$ we have to deal with an infinite set
of complex curves $E_N(\nu_h;n_h,\Mybf{\ell})$ labelled by
$\Mybf{\ell}-$integers. For $\nu_h$ real, they define the energy of ``physical''
$N-$reggeon states, whose wave functions diagonalize the integrals of motion
$\Mybf{q}$ and satisfy the normalization condition \re{norm}. Going over to
complex $\nu_h$, one preserves the former condition and relaxes the latter one.

For $\nu_h$ real, the integrals of motion in the holomorphic and antiholomorphic
sector, $q_k$ and $\bar q_k$, respectively, are conjugated to each other with
respect to the $SL(2,\mathbb{C})$ scalar product \re{norm} so that $\bar
q_k=q_k^*$ (with $k=2,\ldots,N$). Going over to complex $\nu_h$, one finds that
$\bar q_k\neq q_k^*$ and, in general, the quantum numbers in the two sectors are
independent on each other. Nevertheless, the $N-$reggeon spectrum contains the
states, whose integrals of motion take either real, or imaginary values on the
real $\nu_h-$axis, that is $\bar q_k(\nu_h;n_h,\Mybf{\ell})=\pm
q_k(\nu_h;n_h,\Mybf{\ell})$. For such states, the same relation between $q_k$ and
$\bar q_k$ also holds for complex $\nu_h$ but the charges take complex values.

Due to complete integrability, the energy of the $N-$reggeon state is a function
of the integrals of motion $E_N=E_N(q_2,\bar q_2,\ldots,q_N,\bar q_N)$ with
$q_k=q_k(\nu_h;n_h,\Mybf{\ell})$ and similar for $\bar q_k$. For $\nu_h$ real,
the energy spectrum is described by infinite set of smooth real functions
$E_N(\nu_h;n_h,\Mybf{\ell})$ labelled by integers $n_h$ and $\Mybf{\ell}$. The
minimal value of $E_N$ on the real $\nu_h-$axis determines the position of the
dominant Regge singularity $j_N$, Eq.~\re{sq-root}. For $\nu_h$ complex, one
expects that, similar to the $N=2$ case, the complex curve
$E_N(\nu_h;n_h,\Mybf{\ell})$ has an infinite number of poles located at
$i\nu_h=\pm(n-1)/2$ with $n\ge N+n_h$. As we will show below, this turns out to
be the case but in comparison with the $N=2$ case one encounters a novel
phenomenon. For $N\ge 3$, in addition to the poles, the complex curve
$E_N(\nu_h;n_h,\Mybf{\ell})$ contains (an infinite number of) square-root
branching points on the complex $\nu_h-$plane. Therefore, $E_N$ is a
\textit{multi-valued} function on the complex $\nu_h-$plane and its different
branches are enumerated by integer $n_h$ and $\Mybf{\ell}$. This property is in
agreement with the previous findings of Refs.~\cite{LRS,B}.

To construct the complex curve $E_N(\nu_h;n_h,\Mybf{\ell})$, we apply the
approach developed in Ref.~\cite{DKKM}. Namely, we start with the expression for
the energy $E_N$ obtained there for real $\nu_h$ and analytically continue it to
complex $\nu_h$ following the procedure described above. In this way, one gets
\be
E_N=\frac14\left[\varepsilon(h,q)+\varepsilon(h,-q)+\lr{\varepsilon(1-\bar
h^*,\bar q^*)}^* +\lr{\varepsilon(1-\bar h^*,-\bar q^*)}^*\right],
\label{E-epsilon}
\ee
where the $SL(2,\mathbb{C})$ spins $h$ and $\bar h$ are given by \re{h} and the
notation was introduced for the quantum numbers in the two sectors, $q=\{q_k\}$
and $\bar q=\{\bar q_k\}$ (with $k=2,\ldots,N$). In similar manner,
$-q\equiv\{(-1)^kq_k\}$, $\bar q^*\equiv\{\bar q^*_k\}$ and $-\bar
q^*\equiv\{(-1)^k\bar q^*_k\}$, so that $E_N(-q,-\bar q)=E_N(q,\bar q)$. For real
$\nu_h$ one has $1-\bar h^*=h$ and $\bar q^*=q$, so that \re{E-epsilon} produces
real values for $E_N(q,\bar q)$.  The function $\varepsilon(h,q)$ entering
\re{E-epsilon} is defined for arbitrary complex $h$ and $q$ as
\be
\varepsilon(h,q)=i\frac{d}{d\epsilon}\ln \left[\epsilon^N
Q(i+\epsilon;h,q)\right]\bigg|_{\epsilon=0}\,,
\label{e}
\ee
where $Q(u;h,q)$, the so-called chiral Baxter block, has the following integral
representation
\be
Q(u;h,q)=
\int_0^1 {dz}\, z^{iu-1} Q_1(z)\,.
\label{ans}
\ee
Here the function $Q_1(z)$ is the solution to the $N$th order Fuchsian
differential equation
\be
\left[(z\partial_z)^N z+(z\partial_z)^N z^{-1}-2(z\partial_z)^N
-\sum_{k=2}^N i^k q_k(z\partial_z)^{N-k} \right] Q_1(z)=0\,,
\label{diff}
\ee
with the prescribed asymptotic behaviour at regular singular point $z=1$,
$Q_1(z)\sim (1-z)^{-h-1}$. Since $Q_1(z) \sim \ln^{N-1} z$ for $z\to 0$, the
chiral block $Q(u;h,q)$ defined in \re{ans} is a meromorphic function of $u$ with
the $N$th order poles located at $u=ik$ for $k$ positive integer. It is
convenient to normalize $Q_1(z)$ in \re{ans} in such a way that the residue at
the $N$th order pole $u=i$ equals unity. Together with \re{e} this leads to
\be
Q(i+\epsilon;h,q)=\frac1{\epsilon^N}-i\frac{\varepsilon(h,q)}{\epsilon^{N-1}}+
\mathcal{O}\left(\frac1{\epsilon^{N-2}}\right)\,.
\label{norm1}
\ee
To evaluate $\varepsilon(h,q)$, one has to solve the differential equation
\re{diff} and match the resulting expression for $Q(u;h,q)$, Eq.~\re{ans}, into
\re{norm1}. This allows one to determine \re{E-epsilon} for arbitrary complex $q$
and $\bar q$.

The quantization conditions for the integrals of motion $\Mybf{q}=(q_2,\bar
q_2,\ldots,q_N,\bar q_N)$ follow from the requirement for their common
eigenfunctions $\Psi_{\mybf{q}}(\{\vec z\};\vec z_0)$ to be single-valued
functions on the $\vec z-$plane. These conditions take the form~\cite{DKKM}
%
\be
Q(i+\epsilon;h,q)Q(i-\epsilon;\bar h,-\bar
q)-Q(i+\epsilon;1-h,q)Q(i-\epsilon;1-\bar h,-\bar q)=\mathcal{O}(\epsilon^0)\,.
\label{qc1}
\ee
Substituting \re{norm1} into this relation, one finds that the l.h.s.\ scales as
$1/\epsilon^{2N-1}$ for $\epsilon\to 0$. Requirement for the poles at
$\epsilon=0$ to vanish leads to the overdetermined system of $2N-1$ equations for
the $2(N-2)$ charges $q_k$ and $\bar q_k$ ($k=3,\ldots,N$). The two remaining
charges are given by $q_2=-h(h-1)$ and $\bar q_2=-\bar h(\bar h-1)$. Solving this
system, one obtains the quantized values of the integrals of motion and verifies
that the obtained expressions satisfy three consistency conditions.

Applying Eqs.~\re{E-epsilon}\,--\,\re{qc1} one can calculate
the energy of the $N-$reggeon state for arbitrary complex $\nu_h$. For $\nu_h$
real, they lead to the results for the energy $E_N(\nu_h;n_h,\Mybf{\ell})$
obtained before in Ref.~\cite{DKKM}. For $\nu_h$ complex, they define analytical
continuation of the function $E_N(\nu_h;n_h,\Mybf{\ell})$. At $N=2$ the
differential equation \re{diff} can be solved in terms of Legendre functions and
the resulting expression for the energy \re{E-epsilon} coincides with the known
expression \re{E2-anal}. For $N\ge 3$ the analysis is more involved since
\re{diff} cannot be solved exactly and one has to rely on a power-series
solutions to \re{diff} described at length in Ref.~\cite{DKKM}. In the next
section, we summarize the results for $N=3,4,5,6$ reggeon states.

\section{Spectral surface}

\begin{figure}[th]
\vspace*{3mm} \centerline{{\epsfysize6cm \epsfbox{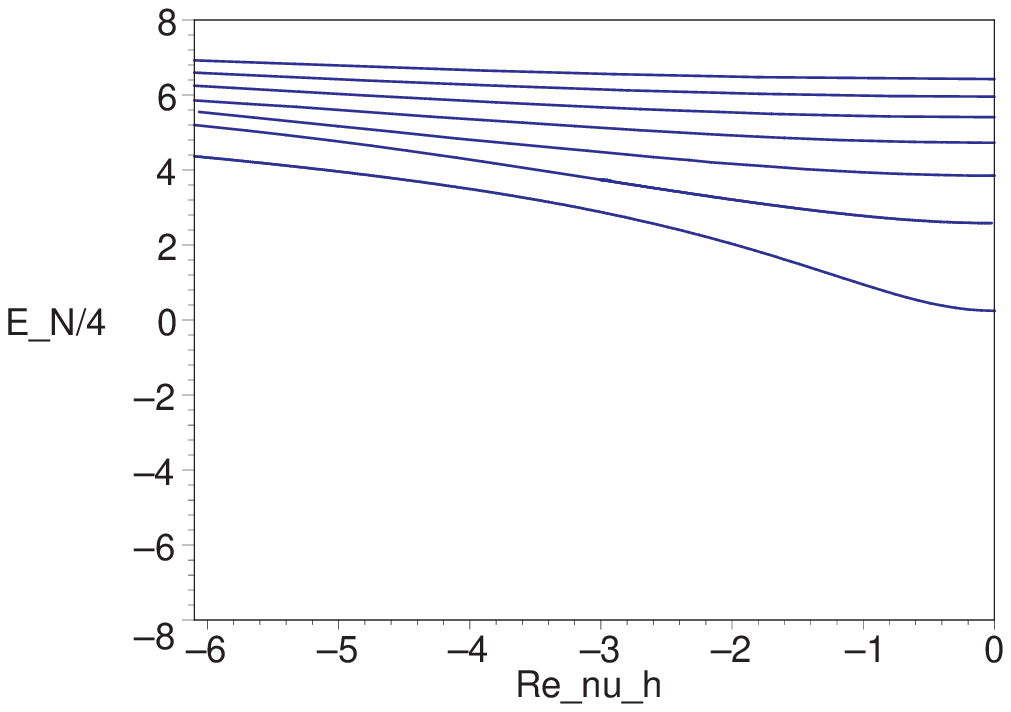}}{\epsfysize6cm
\epsfbox{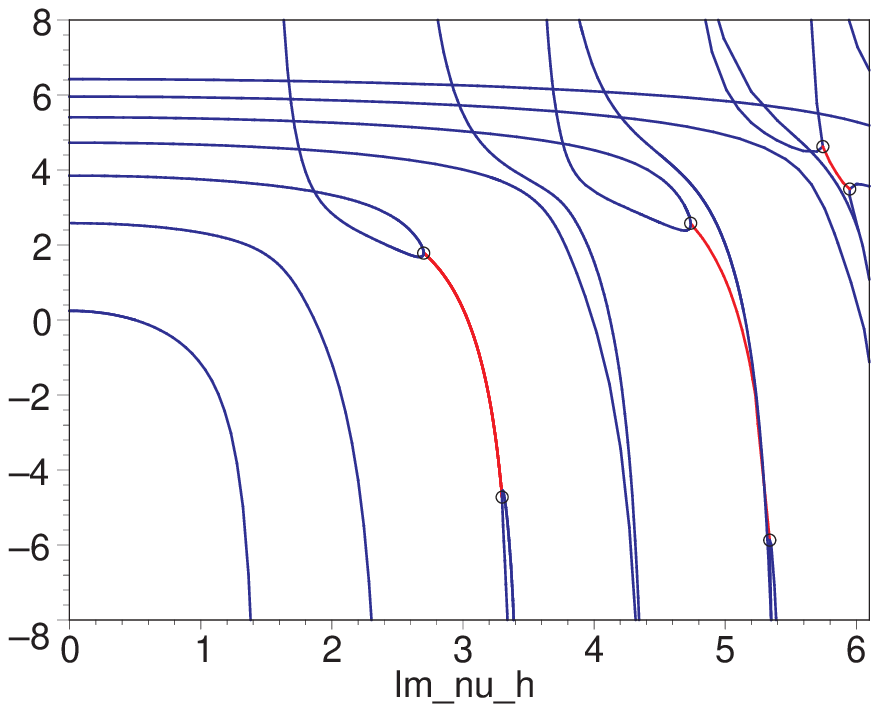}}}
%
%
%
\caption[]{The energy spectrum  of the $N=3$ reggeon states $E_3(\nu_h;n_h,\Mybf{\ell})$
for $n_h=0$ and $\Mybf{\ell}=(0,\ell_2)$, with $\ell_2=2,\,4,\,\ldots,14$ from
the bottom to the top (on the left). Analytical continuation of the energy along
the imaginary $\nu_h-$axis  (on the right). The branching points are indicated by
open circles. The lines connecting the branching points represent $\Re E_3$.}
\label{Fig-E-comp}
\end{figure}

At $N=2$, the energy spectrum $E_2(\nu_h;n_h)$ is described on the complex
$\nu_h-$plane by the family of {\it meromorphic\/} functions \re{E2-anal}
labelled by a nonnegative integer $n_h$. For $N\ge 3$, an analytical expression
for the energy is not available but the value of $E_N$ can be calculated for
arbitrary complex $\nu_h$ using Eqs.~\re{E-epsilon}--\re{qc1} as follows. Let us
consider an arbitrary contour on the complex $\nu_h-$plane that starts on the
real axis and terminates at some complex $\nu_h$. For given real $\nu_h$, the
energy $E_N(\nu_h;n_h,\Mybf{\ell})$ takes a discrete set of real ``physical''
values labelled by integer $n_h$ and $\Mybf{\ell}=(\ell_1,\ldots,\ell_{2N-4})$
(see the left panel in Fig.~\ref{Fig-E-comp}) \cite{DKKM1,DKKM}. Calculating the
energy along this contour point-by-point from Eqs.~\re{E-epsilon}\,--\,\re{qc1},
one analytically continues the function $E_N(\nu_h;n_h,\Mybf{\ell})$ to complex
$\nu_h$.

This procedure allows us to determine the global properties of the complex curve
$E_N(\nu_h;n_h,\Mybf{\ell})$ on the complex $\nu_h-$plane. If $E_N$ were a
single-valued function of complex $\nu_h$, it would resume its original value
after going around arbitrary closed contour on the $\nu_h-$plane. This is the
case at $N=2$, Eq.~\re{E2-anal}, whereas for $N\ge 3$ one finds that
$E_N(\nu_h;n_h,\Mybf{\ell})$ is a \textit{multi-valued} function of $\nu_h$.
Namely, $E_N$ has (an infinite number of) branching points on the complex plane,
$\nu_h=\nu_{{\rm br},k}$, such that $E_N$ changes its value after encircling
these points. We found however that $E_N$ resumes its value if one encircles the
branching point twice. This implies that $E_N$ has square-root cuts
\be
E_N^\pm
\sim a_k \pm b_k \sqrt{\nu_{{\rm br},k}-\nu_h}\,,
\label{cuts}
\ee
where $E_N^\pm$ defines the energy on the upper and lower edges of the cut,
respectively. Aside of the branch cuts, $E_N$ has an infinite number of poles. In
agreement with our expectations, Eq.~\re{expect}, they are located along
imaginary axis at $i\nu_h=\pm (n-1)/2$ with $n\ge N+n_h$.

Thus, for $N\ge 3$ the energy $E_N(\nu_h;n_h,\Mybf{\ell})$ is a meromorphic
function on the complex $\nu_h-$plane with the square-root cuts running between
the branching points $\nu_{{\rm br},k}$. Another way to represent the family of
multivalued functions $E_N(\nu_h;n_h,\Mybf{\ell})$ is to sew together their
branches along the square-root cuts, Eq.~\re{cuts}, and define $E_N$ as a
single-valued meromorphic function on the resulting Riemann surface. Following
\cite{KN}, we shall call it the \textit{spectral surface}.  Its topology depends
on the number of reggeons $N$. Different sheets of this surface can be enumerated
by integer $n_h$ and $\Mybf{\ell}$.

Let us consider a particular $(n_h,\Mybf{\ell})-$th sheet of the spectral surface
lying over the $\nu_h-$plane and suppose that it is sewed with
$(n_h',{\Mybf{\ell}\,}')-$th sheet along the square-root cut that starts at the
branching point $\nu_{\rm br}$. For real $\nu_h$, the value of function $E_N$ on
these two sheets defines the ``physical'' energy of two $N-$reggeon states,
$E_N(\nu_h;n_h,\Mybf{\ell})$ and $E_N(\nu_h;n_h',{\Mybf{\ell}\,}')$,
respectively. If one analytically continues these functions along the contour
that starts at real $\nu_h$ and terminates at the branching point,
$\nu_h=\nu_{{\rm br}}$, then at the vicinity of the branching point the functions
behave as \re{cuts} leading to
\be
E_N(\nu_{{\rm br}};n_h,\Mybf{\ell})=E_N(\nu_{{\rm br}};n_h',{\Mybf{\ell}\,}')\,.
\label{collision}
\ee
Thus, in a complete analogy with quantum mechanics~\cite{BW}, the branching cuts
\re{cuts} arise due to collision of the energy levels at some complex
$\nu_h=\nu_{{\rm br}}$ away from the real $\nu_h-$axis. Since the wave functions
of the two states have to coincide at the branching point~\cite{BW}, they have to
have the same two-dimensional Lorentz spin, $n_h=n_h'$, and possess the same
quantum numbers (quasimomentum, C-parity, Bose symmetry, etc) leading to
additional selection rules for $\Mybf{\ell}$ and ${\Mybf{\ell}\,}'$. This implies
that the spectral surface can not be simply connected and it should consist of
(infinite number of) disconnected components enumerated by nonnegative integer
$n_h$ and the quantum numbers just mentioned.

\psfrag{E_N/4}[cc][cc]{$E_3$} \psfrag{Im_nu_h}[cc][bc]{$\mbox{\large$i\nu_h$}$}
\psfrag{Re_nu_h}[cc][bc]{$\mbox{\large$\nu_h$}$}

To illustrate our findings, let us consider the $N=3$ reggeon states with the
Lorentz spin $n_h=0$. Their spectrum is specified by two integers
$\Mybf{\ell}=(\ell_1,\ell_2)$. Assigning the values of these integers to
different energy levels, we follow the convention adopted in Ref.~\cite{DKKM}.
The energy $E_3(\nu_h;0,\Mybf{\ell})$ is a smooth even function on the real
$\nu_h-$axis (see Fig.~\ref{Fig-E-comp} on the left). It approaches the minimal
value $\min_{\nu_h\mybf{\ell}} E_3=.24717$ at $\nu_h=0$ along the ``trajectory''
with $\ell_1=0$ and $\ell_2=2$. The charges $q_3(\nu_h)$ and $\bar q_3(\nu_h)$
take pure imaginary values along this trajectory, so that $\bar q_3+q_3=0$. After
analytical continuation, the same relation holds for arbitrary complex $\nu_h$
including the branching point \re{collision}. Therefore, the energy level with
$\ell_1=0$ and $\ell_2=2$ could only collide with the levels for which $\bar
q_3+q_3=0$. As was shown in \cite{DKKM}, the latter condition is satisfied for
$\ell_1=0$ and $\ell_2=$ even positive. Examples of such energy levels with
$\ell_2=2,4,\!...,14$ are shown in Fig.~\ref{Fig-E-comp}. They define a connected
component of the $N=3$ spectral surface.

To elucidate analytical properties of the energy, we used
Eqs.~\re{E-epsilon}--\re{qc1} to analytically continue eight functions
$E_3(\nu_h;0,\Mybf{\ell}=(0,\ell_2))$ with $\ell_2=2,4,\!...,14$. These functions
define eight different branches of the complex curve $E_3(\nu_h)$ which can be
represented as a two-dimensional surface in $\mathbb{C}^2=\mathbb{R}^4$  space
with the $(\nu_h,E_3)-$coordinates. The slices of this surface along $\Im\nu_h=0$
and $\Re\nu_h=0$ hyperplanes are shown in Fig.~\ref{Fig-E-comp} on the left and
the right panels, respectively. We observe that, firstly, the energy has poles at
$i\nu_{\rm pole}=3/2,\, 5/2,\, 7/2,\,9/2,\,11/2$ and, secondly, different energy
levels collide at
$i\nu_{\rm br}=
    2.70
,\, 3.29
,\, 4.73
,\, 5.34
,\, 5.74
,\, 5.94
$.
At the vicinity of the branching point the energies of two colliding levels
behave as $E_3^\pm$, Eq.~\re{cuts}. As expected, the poles of $E_3$ are located
along the imaginary $\nu_h-$axis at $i\nu_h=(n-1)/2$ with $n=4,\,6,\!...$. This
is not the case, however, for the branching points. It turns out that some of the
branching points are located away from the imaginary $\nu_h-$axis and, therefore,
they can not be seen in Fig.~\ref{Fig-E-comp}. For instance, the branches with
$\ell_2=2$ and $\ell_2=4$ collide at $i\nu_{\rm br}=1.723+i\,.248$. In other
words, the ``ground state'' branch, $\ell_2=2$, is sewed with the $\ell_2=4$
branch which in its turn is sewed with the $\ell_2=6$ branch and so on. Thus,
going around the contour on the complex $\nu_h-$plane which starts at some real
$\nu_h$, encircles the branching points and returns to the starting point one can
reconstruct the energy spectrum of the $N=3$ reggeon states with $\ell_1=0$ and
$\ell_2=$ even positive.

The fact that the branching points $\nu_{\rm br}$ take complex irrational values
implies that the contribution of the corresponding square-root cuts to the
integral \re{F-fin} scales as $1/[Q^{1+2i\nu_{\rm br}}\ln ^{3/2} Q]$ and,
therefore, it breaks the OPE expansion \re{moments}. Although such corrections
are present in \re{F-fin} for given $n_h$ and $\Mybf{\ell}$, they cancel against
each other in the sum over all states. To see this let us examine two terms in
the sum \re{F-fin} corresponding to the energy levels colliding at
$\nu_h=\nu_{\rm br}$. Each of them is associated with a particular sheet of the
spectral surface. Denoting the energy on the upper $(+)$ and the lower $(-)$
edges of the square-root cut on these two sheets as $E_1^\pm(\nu_h)$ and
$E_2^\pm(\nu_h)$ one finds that $E_1^\pm(\nu_h)=E_2^\mp(\nu_h)$. As a
consequence, the contour integrals around the same cut on the two sheets differ
by a sign and their sum equals zero. Thus, the contribution of the cuts to
\re{F-fin} cancels completely in the sum over \textit{all} $N=3$ reggeon states
belonging to the same spectral surface. At the same time, if one had retained in
\re{F-fin} only contribution of the state with the minimal energy $E_{3}^{\rm
(gr)}$ -- the one defining the leading Regge singularity \re{sq-root}, the OPE
expansion would have been broken. To restore the OPE one has to keep in
\re{F-fin} the contribution of subleading Regge singularities.

As another important example, we consider analytical continuation of the $N=3$
reggeon state with the Lorentz spin $n_h=1$, the so-called descendant
state~\cite{DKKM}. This state has $q_3(\nu_h)=0$ on the real $\nu_h-$axis and its
energy coincides with the energy of the $N=2$ state, $E_{3,{\rm
d}}=E_2(\nu_h;1)$, Eq.~\re{E2-anal}. Both relations survive the analytical
continuation and hold true for arbitrary complex $\nu_h$. Thus, in distinction
with the previous case, the energy of this state $E_{3,{\rm d}}(\nu_h)$ is a
single-valued, meromorphic function on the complex $\nu_h-$plane. At $\nu_h=0$ it
takes the value $E_3=E_2(0;1)=0$ which defines the ``physical'' ground state
energy for the system of $N=3$ reggeons~\cite{BLV}. Going over to the imaginary
$\nu_h-$axis, one finds that $E_3$ has poles in the lower half-plane at $i\nu_h=
1,\, 2,\,\!...$

We would like to stress that the $N=3$ states with $n_h=0$ and $n_h=1$ described
above define two different components of the unifying $N=3$ spectral surface. The
reason why we have selected them is that they contain two special states
$E_3(\nu_h=0;n_h=1)=0$ and $E_3(\nu_h=0;n_h=0)=.24717$, which are currently
considered as solutions for the Odderon state in QCD with the intercept
$j_\mathbb{O}=1-\bar\alpha_s E_3$~\cite{JW,BLV}. As was shown in Sect.~2, the
poles of the energy in the lower half-plane $\Im\nu_h<0$, induce the contribution
to the OPE expansion \re{moments} of the twist $n=h+\bar h=1+2i\nu_h$. The
minimal twist $n_{\rm min}$ corresponds to the pole closest to the origin. As
follows from our analysis, $n_{\rm min}=3$ for the descendant state with $n_h=1$
\footnote{We disagree on this point with Ref.~\cite{DL}, in which it was claimed
that $n_{\rm min}=4$ for the Odderon state with the intercept $1$.} and $n_{\rm
min}=4$ for the state with $n_h=0$. Thus, the leading twist contribution of the
two above mentioned Odderon states with the intercepts $j_\mathbb{O}=1$ and
$j_\mathbb{O}<1$ scales at large $Q^2$ as $\sim 1/Q^3$ and $\sim 1/Q^4$,
respectively.

\section{Anomalous dimensions}

To calculate the anomalous dimensions \re{gamma-pole-exp}, one has to work out
the Laurent expansion of the energy around its poles, Eq.~\re{E-pole-exp}. We
recall that for the $N-$reggeon state the twist$-n$ contribution comes from poles
located at $i\nu_h=(n-1)/2$ with $n\ge N+n_h$, or equivalently $h=(n+n_h)/2$ and
$\bar h=(n-n_h)/2$.

Let us first consider the $N=3$ descendant state, $n_h=1$. For complex $\nu_h$
its energy, $E_{3,\,\rm d}$, coincides with $E_2(\nu_h,1)$ defined in
\re{E2-anal}. The closest to the origin pole is located at $i\nu_h=1$ and the
corresponding $SL(2,\mathbb{C})$ spins \re{h} are given by $h=1+i\nu_h=2$ and
$\bar h=i\nu_h=1$ so that the twist equals $n=h+\bar h=3$. Using \re{E2-anal},
one finds the expansion of the energy $E_{3,\,\rm d}=E_{3,\,\rm d}(h+\epsilon)$
for $\epsilon\to 0$ as
\be
E_{3,\rm d}(2+\epsilon)=\frac1{\epsilon}+1-\epsilon -\left( 2\zeta(3) -1
\right) {\epsilon}^{2}+\ldots
\label{E3d}
\ee
Applying Eqs.~\re{E-pole-exp} and \re{gamma-pole-exp}, one finds from \re{E3d}
the twist-3 anomalous dimension corresponding to the Odderon state with the
intercept $j_\mathbb{O}=1$ as ($\omega=j-1$)
\be
\gamma_{3}^{(N=3)}(j) = \frac{\bar\alpha_s}{\omega}-\lr{\frac{\bar\alpha_s}{\omega}}^2
+(2\zeta(3)+1)\lr{\frac{\bar\alpha_s}{\omega}}^4+\ldots\,,
\label{gamma3}
\ee
where the subscript and the superscript indicate the twist and the number of
reggeons entering the state, respectively.

Let us now consider the $N=3$ states with $n_h=0$. Their analytical properties
are encoded in the spectral surface shown in Fig.~\ref{Fig-E-comp}. Applying
Eqs.~\re{E-epsilon}--\re{qc1}, we calculate the energy at the vicinity of the
poles at $h=1/2+i\nu_h=2,\,3,\,4,\,5$ and obtain the following expressions for
$E_3=E_3(h+\epsilon)$
\ba
E_3(2+\epsilon)&=& {\epsilon}^{-1} + \frac12 -\frac12\ \epsilon + 1.7021\,
\epsilon^2+\ldots
\nonumber\\
E_3(3+\epsilon)&=& {2}\,{\epsilon}^{-1} + {\frac {15}{8}} - 1.6172\ \epsilon +
 0.719\ {\epsilon}^{2}+\ldots
\nonumber \\
E_3^{\rm(a)}(4+\epsilon)&=& {\epsilon}^{-1} + \frac{11}{12} - 0.6806\
\epsilon - 1.966\ \epsilon^2+\ldots
\nonumber \\
E_3^{\rm(b)}(4+\epsilon)&=& {2}\,{\epsilon}^{-1}+\frac{15}4-3.2187\ \epsilon
 + 3.430\ \epsilon^2+\ldots
\nonumber \\
E_3^{\rm(a)}(5+\epsilon)&=& 2\,{\epsilon}^{-1} + \frac{125}{48} - 2.0687\
\epsilon + 1.047\ \epsilon^2+\ldots
\nonumber \\
E_3^{\rm(b)}(5+\epsilon)&=& {2}\,{\epsilon}^{-1} + \frac{53}{12} - 2.4225\
\epsilon + 0.247\ {\epsilon}^{2}+\ldots
\label{res-N3}
\ea
Here ellipses denote $\mathcal{O}(\epsilon^3)$ terms and the additional
superscript was introduced to distinguish between different branches. We recall
that all states in \re{res-N3} have the same Lorentz spin $n_h=0$, so that $\bar
h=h=1/2+i\nu_h$ and the corresponding twist equals $n=2h\ge 4$. Applying
Eqs.~\re{E-pole-exp} and \re{gamma-pole-exp}, one finds from the first relation
in \re{res-N3} that the leading, twist-4 anomalous dimension corresponding to the
Odderon state with the intercept $j_\mathbb{O}<1$ is given by
\be
\gamma_4^{(N=3)}(j) =
\frac{\bar\alpha_s}{\omega}-\frac12\lr{\frac{\bar\alpha_s}{\omega}}^2
-\frac14\lr{\frac{\bar\alpha_s}{\omega}}^3-1.0771
\lr{\frac{\bar\alpha_s}{\omega}}^4+\ldots
\label{gamma3-1}
\ee
The remaining relations in \re{res-N3} lead to similar expressions for
higher-twist anomalous dimensions. To save space we do not present them here.
The following comments are in order.

The first few terms of the Laurent expansion in \re{res-N3} can be calculated
exactly. The reason for this is that solutions to the differential equation
\re{diff} can be expanded in powers of $\epsilon$ and the first few terms can be
obtained in a closed form. The coefficient in front of $1/\epsilon$ in
\re{res-N3} equals either $1$ (only for $h$ even), or $2$. In the latter case,
writing the energy as $E_3(h+\epsilon) =2\epsilon^{-1} + \gamma(h)
+\mathcal{O}(\epsilon)$ one finds that finite $\mathcal{O}(\epsilon^0)-$terms
have the following remarkable property. It turns out that $\gamma(4)=15/4$ and
$\gamma(5)=53/12$ coincide with the energy of the Heisenberg $SL(2,\mathbb{R})$
magnet model of spin $s=1$. Most importantly, the energy spectrum of this model
determines the anomalous dimensions of local composite \textit{three-quark}
(baryonic) operators of helicity$-3/2$~\cite{BDKM,AB}. For such operators,
integer $(h-3)$ counts the number of covariant derivatives and $\gamma(h)$
defines their anomalous dimension. Similar relation holds for higher $N$ between
the energy of the $N-$reggeon states ($SL(2,\mathbb{C})$ Heisenberg spin magnet)
and anomalous dimensions of $N-$particle operators ($SL(2,\mathbb{R})$ Heisenberg
spin magnet). Its origin will be discussed in a forthcoming publication.

Going over to the $N\ge 4$ case, we shall concentrate on the $N-$reggeon states
providing the contribution to the OPE expansion \re{moments} of the minimal twist
$n_{\rm min}$. As before, we shall denote the corresponding anomalous dimension
as $\gamma_{\,{n_{\rm min}}}^{_{(N)}}(j)$.

For $N=$~\textit{even} we find that the minimal twist equals the number of
reggeons involved $n_{\rm min}=N$. It corresponds to the pole of the complex
curve $E_N$ located at $i\nu_h=(N-1)/2$ and $n_h=0$, or equivalently $h=\bar
h=N/2$. This pole is situated on the same spectral surface as the ``physical''
ground $N-$reggeon state, that is the state with the minimal energy for real
$\nu_h$. Its energy, $E_{N}^{\rm (gr)}=\min_{\nu_h n_h\mybf{\ell}}
E_N(\nu_h;n_h,\Mybf{\ell})$, defines the intercept of the $N-$reggeon states in
the Pomeron sector~\cite{DKKM1,DKKM}: $E_{4}^{\rm (gr)}=-.67416$ and $E_{6}^{\rm
(gr)}=-.39458$. The expansion of the energy $E_N(h+\epsilon)$ around $\epsilon=0$
looks like
\ba
E_4(2+\epsilon)&=& \frac2{\epsilon} + 1 - \frac12\,\epsilon -
1.2021\,{\epsilon}^{ 2} + \ldots
\nonumber
\\
E_6(3+\epsilon)&=& \frac4{\epsilon} + \frac32
-
\frac7{16}
\,\epsilon-0.238\,\epsilon^2+\ldots
\label{E46-pol}
\ea
Applying Eqs.~\re{E-pole-exp} and \re{gamma-pole-exp}, one obtains the following
expressions for the leading, twist$-N$ anomalous dimension of the $N-$reggeon
states
\ba
\gamma_4^{(N=4)}(j)&=&2\frac{\bar\alpha_s}{\omega}+
2\lr{\frac{\bar\alpha_s}{\omega}}^2
-13.6168\lr{\frac{\bar\alpha_s}{\omega}}^4+\ldots \nonumber
\\
\gamma_6^{(N=6)}(j)&=&4\,\frac{\bar\alpha_s}{\omega}+
6\,\lr{\frac{\bar\alpha_s}{\omega}}^2 +2\,\lr{\frac{\bar\alpha_s}{\omega}}^3-
33.23\,\lr{\frac{\bar\alpha_s}{\omega}}^4+\ldots
\label{gamma46}
\ea
with $\omega=j-1$. We observe that higher order corrections to \re{gamma46} are
large indicating that the series have a finite radius of convergence and its
value decreases with $N$. This is in a qualitative agreement with the fact
 that the intercept of the $N-$reggeon states scales at large $N$ as
$|j_N-1|\sim 1/N$~\cite{DKKM}. Writing the $\mathcal{O}(\bar\alpha_s)-$correction
to \re{gamma46} as $\gamma_{2n}(\omega) =\varepsilon_{n}{\bar\alpha_s}/{\omega}$
(with $n=2,3$), one finds that $\varepsilon_{n}$ verifies the condition
$\varepsilon_{n}\le 2(n-1)$ established in \cite{S}. In addition,
$\gamma_{2n}(\omega) < n\gamma_2(\omega/n)$ which means that the anomalous
dimension of $N-$reggeon states is subleading in the multi-color limit as
compared with the anomalous dimension of $N/2$ BFKL Pomerons~\cite{LRS,B}.

For $N=$~\textit{odd}, similar to the $N=3$ case, we consider separately the
sectors with the Lorentz spin $n_h=0$ and $n_h=1$. In the first case, the minimal
twist equals $n_{\rm min}=(N+1)$ and it corresponds to the pole of the energy at
$i\nu_h=N/2$ and $n_h=0$, or equivalently $h=\bar h=(N+1)/2$. For instance, at
$N=5$ the expansion of the energy $E_N(h+\epsilon)$ and the anomalous dimension
look like
\ba
E_5(3+\epsilon) &=& \frac3{\epsilon} +  \frac76  - 
\frac{101}{216}\ \epsilon - 0.1136 \
\epsilon^2 +\ldots\,.
\label{E5-pol}
\\
\gamma_6^{(N=5)}(j) &=& 3\,\frac{\bar\alpha_s}{\omega}
+\frac72\lr{\frac{\bar\alpha_s}{\omega}}^2-
\frac18
\lr{\frac{\bar\alpha_s}{\omega}}^3
-13.032\lr{\frac{\bar\alpha_s}{\omega}}^4+\ldots\,.
\nonumber
\ea
The pole \re{E5-pol} is located on the same spectral surface as the physical
ground state in this sector, $E^{\rm (gr)}_5(n_h=0)=.12751$. This state has the
quantum numbers of the Odderon. It has the intercept $1-\bar\alpha_s E^{\rm
(br)}_5$ which is smaller then 1 and its leading twist equals $6$.

In the second case, for $N=$~\textit{odd} and $n_h=1$, the minimal twist is
smaller $n_{\rm min}=N$. It corresponds to the pole of the energy $E_N$ at
$i\nu_h=(N-1)/2$, or equivalently $h=(N+1)/2$ and $\bar h=(N-1)/2$. At $N=5$ the
expansion of the energy $E_N(h+\epsilon)$ looks like
\ba
&&E_{5,\rm d}(3+\epsilon)=
\frac{3+\sqrt{5}}{2\epsilon} + 1.36180 - 0.4349 \ \epsilon - 0.315\ \epsilon^2 +\ldots
\label{pole-des-5}
\\
&&\gamma_5^{(N=5)}(j)= \frac{3+\sqrt{5}}2\frac{\bar\alpha_s}{\omega}
+3.56524\lr{\frac{\bar\alpha_s}{\omega}}^2
+1.8743\lr{\frac{\bar\alpha_s}{\omega}}^3
-11.219\lr{\frac{\bar\alpha_s}{\omega}}^4+\ldots \nonumber
\ea
As in the $N=3$ case, this reggeon state is descendant~\cite{DKKM}, that is its
energy equals the energy of the $N=4$ state with $n_h=1$. The pole
\re{pole-des-5} belongs to the same spectral surface as the $N=5$ ground state,
$E^{_{\rm (gr)}}_5(n_h=1)=0$, which has the quantum numbers of the Odderon and
intercept equal to unity. The second relation in \re{pole-des-5} defines the
leading, twist$-5$ anomalous dimension of this state.


\section{Conclusions}

In this paper, we performed the OPE analysis of the contribution of $N-$reggeon
compound states to the moments of the structure function $\widetilde F(j,Q^2)$
for $j\to 1$ and calculated the anomalous dimensions of the leading twist
contribution in multi-color QCD. To this end, we analytically continued the
energy of the $N-$reggeon states $E_N$ from the ``physical'' region of parameters
(real $\nu_h-$axis) to complex $\nu_h$ and established the relation between the
anomalous dimensions for $j\to 1$ and the Laurent expansion of $E_N$ around its
poles. At $N=2$ the energy is a meromorphic function on the complex
$\nu_h-$plane~\cite{L85} while for $N\ge 3$ analytical properties of the energy
$E_N$ are changed dramatically. Namely, we found that, in agreement with previous
findings \cite{LRS,B}, the energy $E_N$ is a multi-valued function on the complex
$\nu_h-$plane. The reason for this is that different energy levels in the
spectrum of the $N-$reggeon states collide after analytical continuation to
complex $\nu_h$ and, as a consequence, their energies develop square-root cuts.
Due to nonvanishing contribution of the cuts, each energy level (the sheet of the
spectral surface) breaks the twist expansion of $\widetilde F(j,Q^2)$, but it is
restored in the sum over all $N-$reggeon states.

To summarize our main results, we found that the leading contribution of the
$N-$reggeon states to the moments \re{moments} has the twist $N$. For even and
odd $N$ it comes from the reggeon states with the two-dimensional angular
momentum $n_h=0$ and $n_h=1$, respectively. For $N=3,4,5,6$ the corresponding
anomalous dimensions $\gamma_N^{_{(N)}}$ are given by Eqs.~\re{gamma3},
\re{gamma46}, \re{pole-des-5}. The explicit form of the underlying twist$-N$
operators need to be found. We demonstrated that two solutions for the Odderon
states with the intercepts $j_\mathbb{O}=1$ and $j_\mathbb{O}<1$, recently
discussed in the literature, have the twist $3$ and $4$, respectively. Their
anomalous dimensions are defined in Eqs.~\re{gamma3} and \re{gamma3-1}. Notice
that one-loop corrections to the twist$-N$ anomalous dimensions,
Eqs.~\re{gamma46}--\re{pole-des-5}, have a rather simple form. It would be
interesting to compare these expressions with those obtained by explicit
calculation of the twist$-N$ anomalous dimensions in the double-logarithmic
approximation (see e.g. \cite{S}).

Finally, let us comment on the relation of our results to those obtained in
Ref.~\cite{DL}. In that paper, the anomalous dimension of the $N=3$ and $N=4$
reggeon states have been calculated in the multi-color limit and the expressions
obtained there differ from Eqs.~\re{gamma3} and \re{gamma46}. The reason for
disagreement is the following. The approach proposed in \cite{DL} is based on the
assumption that for $n_h\neq 0$ the energy of the $N-$reggeon states as a
function of conformal spins $h=(1+n_h)/2$ and $\bar h=(1-n_h)/2$ can be obtained
by analytical continuation in $h$ from the energy in the sector with the Lorentz
spin $n_h=0$. As follows from our analysis, this assumption is erroneous. For
$N\ge 3$ the energy $E_N$ is a multi-valued function of conformal spin
(see Fig.~\ref{Fig-E-comp}) and it can only be analytically
continued over the spectral surface described in Sect.~4. The states with
different spins $n_h$ belong to different \textit{disconnected} components of
this surface so that their energies are not related to each other by analytical
continuation. If one ignored nontrivial analytical properties of the energy and
applied the approach of Ref.~\cite{DL}, one would generate spurious states which
do not belong to the physical spectrum of the reggeon states.

\bigskip

\noindent We have greatly benefited from discussions with Ian Kogan (1958 -- 2003).
We would also like to thank A.~Gorsky, S.~Derkachov, A.~Mueller and A.~Turbiner
for helpful conversations. This work was supported in part by the grant
KBN-PB-2-P03B-043-24 (J.K.), the Sofya Kovalevskaya programme of Alexander von
Humboldt Foundation (A.M.) and by the NATO Fellowship (A.M.).

\bb{99}

\bi{T}  T.~Jaroszewicz, 
        Phys.\ Lett.\ B {\bf 116} (1982) 291.

\bi{BFKL}
        E.A.~Kuraev, L.N.~Lipatov and V.S.~Fadin,
        Phys.\ Lett.\ B\textbf{60} (1975) 50; Sov.\ Phys.\ JETP \textbf{44} (1976) 443;
        45 (1977) 199;
        I.I.~Balitski and L.N~Lipatov, Sov.\ J.\ Nucl.\ Phys.\ \textbf{28} (1978) 822.

\bi{Mueller1996} A.~H.~Mueller,
        Phys.\ Lett.\ B {\bf 396} (1997) 251.

\bi{L85}L.~N.~Lipatov, 
        Sov.\ Phys.\ JETP {\bf 63} (1986) 904.

\bi{DL} H.J.~De~Vega and L.N.~Lipatov, Phys.\ Rev.\ D {\bf 66} (2002) 074013.

\bi{BKP}J.~Bartels, Nucl.\ Phys.\ B {\bf 175} (1980) 365;
        J.~Kwiecinski and M.~Praszalowicz, Phys.\ Lett.\ B {\bf 94} (1980) 413.

\bi{LRS}E.~M.~Levin, M.~G.~Ryskin and A.~G.~Shuvaev,
        Nucl.\ Phys.\ B {\bf 387} (1992) 589.

\bi{B}  J.~Bartels,
        Phys.\ Lett.\ B {\bf 298} (1993) 204;
        Z.\ Phys.\ C {\bf 60} (1993) 471.

\bi{L1} L.N.~Lipatov, JETP Lett. {\bf 59} (1994) 596.

\bi{FK} L.D.~Faddeev and G.P.~Korchemsky,
        Phys.\ Lett.\ B {\bf 342} (1995) 311.

\bi{DKKM1}
        G.~P.~Korchemsky, J.~Kotanski and A.~N.~Manashov,
        Phys.\ Rev.\ Lett.\  {\bf 88} (2002) 122002.

\bi{DKKM}
        S.~E.~Derkachov, G.~P.~Korchemsky and A.~N.~Manashov,
        Nucl.\ Phys.\ B {\bf 617} (2001) 375;
        Nucl.\ Phys.\ B {\bf 661} (2003) 533;
        S.~E.~Derkachov, G.~P.~Korchemsky, J.~Kotanski and A.~N.~Manashov,
        Nucl.\ Phys.\ B {\bf 645} (2002) 237.

\bi{S}  A.~G.~Shuvaev,
        Phys.\ Atom.\ Nucl.\  {\bf 57} (1994) 299; preprints hep-ph/9504341; hep-ph/0310344.

\bi{BW} C.M.~Bender and T.T.~Wu, 
        Phys.\ Rev.\  {\bf 184} (1969) 1231.

\bi{TU} A.V.~Turbiner and A.G.~Ushveridze, Phys.\ Lett.\ \textbf{A126} (1987)
181.

\bi{BT} C.M.~Bender and A.~Turbiner, Phys.\ Lett.\ \textbf{A173} (1993) 442.

\bi{KN} I.V.~Komarov and E.I.~Novikov, Phys.\ Lett.\ \textbf{A186} (1994) 396.

\bi{BLV}J.~Bartels, L.N.~Lipatov and G.P.~Vacca,
        Phys.\ Lett.\ B {\bf 477} (2000) 178.

\bi{JW} R.~Janik and J.~Wosiek, Phys.\ Rev.\ Lett.\ {\bf 79} (1997) 2935;
          {\bf 82} (1999) 1092.

\bi{BDKM}
        V.M. Braun, S.E. Derkachov, G.P. Korchemsky and A.N.~Manashov,
        Nucl.\ Phys.\ B {\bf 553} (1999) 355.

\bi{AB} A.~V.~Belitsky,
        Nucl.\ Phys.\ B {\bf 558} (1999) 259.

\eb

\end{document}